# Generalized scaling of spin qubit coherence in over 12,000 host materials

Shun Kanai[1-4,*], F. Joseph Heremans[5,6], Hosung Seo[7], Gary Wolfowicz[5,6],
Christopher P. Anderson[6,8], Sean E. Sullivan[5], Giulia Galli[5,6,9],
David D. Awschalom[5,6,8], and Hideo Ohno[1,3,4,10,11]

[1] Laboratory for Nanoelectronics and Spintronics, Research Institute of Electrical Communication, Tohoku University, 2-1-1 Katahira, Aoba-ku, Sendai 980-8577, Japan
[2] Division for the Establishment of Frontier Sciences of Organization for Advanced Studies at Tohoku University, Tohoku University, 2-1-1 Katahira, Aoba-ku, Sendai 980-8577, Japan
[3] Center for Science and Innovation in Spintronics, Tohoku University, 2-1-1 Katahira, Aoba-ku, Sendai 980-8577, Japan
[4] Center for Spintronics Research Network, Tohoku University, 2-1-1 Katahira, Aoba-ku, Sendai 980-8577, Japan
[5] Center for Molecular Engineering and Materials Science Division, Argonne National Laboratory, Lemont, IL 60439, USA
[6] Pritzker School of Molecular Engineering, University of Chicago, Chicago, IL 60637, USA
[7] Department of Physics and Department of Energy Systems Research, Ajou University, Suwon, Gyeonggi 16499, Republic of Korea
[8] Department of Physics, University of Chicago, Chicago, IL 60637, USA
[9] Department of Chemistry, University of Chicago, Chicago, IL 60637, USA
[10] Center for Innovative Integrated Electronic Systems, Tohoku University, 468-1 Aramaki Aza Aoba, Aoba-ku,Sendai 980-0845, Japan
[11] WPI-Advanced Institute for Materials Research, Tohoku University, 2-1-1 Katahira, Aoba-ku, Sendai 980-8577, Japan
* e-mail: skanai@tohoku.ac.jp

**Spin defect centers with long quantum coherence times ($T_2$) are key solid-state platforms for a variety of quantum applications[1,2]. Recently, cluster correlation expansion (CCE) techniques have emerged as a powerful tool to simulate the $T_2$ of defect electron spins in these solid-state systems with good accuracy[3–8]. Here, based on CCE, we uncover an algebraic expression for $T_2$ generalized for host compounds with dilute nuclear spin baths, which enables a quantitative and comprehensive materials exploration with a near instantaneous estimate of the coherence. We investigate more than 12,000 host compounds at natural isotopic abundance, and find that silicon carbide (SiC), a prominent widegap semiconductor for quantum applications[1,2,5,8–13], possesses the longest coherence times among widegap non-chalcogenides. In addition, more than 700 chalcogenides are shown to possess a** **longer $T_2$ than SiC. We suggest new potential host compounds with promisingly long $T_2$ up to 47 ms, and pave the way to explore unprecedented functional materials for quantum applications.**

Defect centers have been used to demonstrate a wide range of functionalities, including remote entanglement[15], control of large nuclear spin clusters[16], and quantum sensing of local temperature[17], magnetic field[18], electric field[19], and strain[20]. While these functionalities have been investigated in only a few solid-state systems, new defects and host materials may offer a new range of opportunities. Weber, Koehl, Varley *et al*[1] consolidated the generalized criteria of the preferable properties of the hosts for spin qubit applications[2]: a wide bandgap, small spin-orbit coupling, nuclear spin free lattice, and availability of high-quality single crystal. These criteria led to the identification of SiC as a promising host for qubits, which broadened the field beyond the $NV^-$ center in diamond and uncovered a varied landscape of materials for defect spin qubits with different relative advantages and disadvantages.

For most quantum applications, the key property of interest is the electron spin coherence time, generally defined as $T_2$ by Hahn echo measurement, *i.e.* after refocusing of slow fluctuating noises by a single $\pi$-pulse[10]. The electron spin $T_2$ is limited by the interaction of the spin with its surrounding electric, thermal, and magnetic noise. However, in the absence of additional paramagnetic defects or the spin relaxation time ($T_1$) limitations, in most quantum applications, the electron spin $T_2$ is well predicted by only considering the effect of nuclear spins in the host materials, especially in high quality, wide-bandgap systems at cryogenic temperatures. For a $S$ = 1/2 electron spin interacting with a few $I$ = 1/2 nuclear spins, analytical solutions for the electron spin echo envelope modulation (ESEEM) have existed for half a century[21]. Yet an analytic equation is absent for efficiently predicting $T_2$ of a typical electron spin in a solid-state defect center interacting with several thousand nuclear spins[3,4,7,8,22], which is highly desirable in the wide-range search of new quantum host materials.

Cluster correlation expansion[3,4,7], enables accurate calculations of the $T_2$ of an electron spin interacting with large number of nuclear spins by dividing the spins into small subsets of interacting spin clusters as shown in Fig. 1a. In particular, the pairwise treatment of nuclear spins has been shown to provide an excellent accuracy in simulating the decoherence of spin qubits in several dilute nuclear spin host materials: Bi dopants in silicon, the $NV^-$ center in diamond, and the $VV^0$ center in SiC[6-8]. CCE calculations, however, are still not an easy-to-use prediction scheme, requiring derivations from first principles calculations[23,24] and computationally expensive simulations especially for compounds with $I > 1/2$, making its use for the high throughput search of new qubit host materials limited.

Here, we use CCE to uncover a method to not only to explore over 12,000 host materials for quantum applications and discover candidates with a long electron spin coherence time, but also to provide an easy-to-use $T_2$ prediction scheme. We first investigate how materials with a dilute (< $10^{22}$ $cm^{-3}$) nuclear spin bath comprised of one or multiple nuclear spin species can be decomposed into separate independent baths for each species. We then show that the electron spin $T_2$ of each individual bath is scaled by its nuclear spin $g$-factor value, density, and quantum number regardless of the crystalline structure of the material. This results in a single phenomenological expression for estimating any compounds' $T_2$ without treating the spin Hamiltonian and the time evolution of spins exactly. By calculating $T_2$ for every element in the periodic table and mining materials databases[25,26], we categorize, calculate, and predict many new candidates with a long quantum coherence time. Even though $T_2$ can be limited by interactions other than with the nuclear bath, our results set the fundamental materials limit for spin decoherence when all other sources are eliminated.

To begin, we benchmark our CCE calculations (Methods) on known materials. Figure 1b shows the examples of calculated Hahn-echo signal ($\mathcal{L}(t_{\text{free}})$) using CCE as a function of the free evolution duration ($t_{\text{free}}$) in naturally abundant 4$H$-SiC, diamond, Si as well as typical wide band-gap oxides. We neglect the Fermi contact terms of the hyperfine interaction given the localized electronic nature of deep-level defects and a dilute nuclear spin density in the host. We also adopt the secular approximation for the hyperfine interaction, which holds when $S_z$ is a good quantum number in the presence of a strong $B$, here fixed at 5 T. Within this approximation, the Hamiltonian is reduced into bath Hamiltonians treating only the nuclear spin bath[8], meaning the calculation is mostly agnostic to the spin defect Hamiltonian. This is crucial to allow for wide scale predictions.

$T_2$ is obtained by fitting the calculated $\mathcal{L}(t_{\text{free}})$ with a decay function $e^{-(t_{\text{free}}/T_2)^{\eta}}$, where $\eta$ is a stretching exponent[27]. The envelope of the Hahn-echo signal is critically determined by the dipole-dipole interactions between nuclear spins. Figure 1c shows the $\mathcal{L}(t_{\text{free}})$ of $SiO_2$ (α-quartz) with $B$ = 300 mT, dividing the interactions between baths of homo-, and hetero-nuclear spins in simulation. Heteronuclear spin interactions do not contribute to decoherence in this time range, supporting that the homonuclear spin-spin interaction is the main source of the decoherence due to the decoupling of the heteronuclear spin baths. Generally, when the heteronuclear dipole-dipole interactions are much smaller than the difference of their Zeeman energies, the heteronuclear spin baths are decoupled[22]. We find that even in a worst-case scenario, any heteronuclear spin baths can be decoupled under achievable experimental conditions (Methods).

When heteronuclear spin baths are decoupled, one can simulate a compound's Hahn echo signal by only considering the homonuclear spin baths; $\mathcal{L}(t_{\text{free}})$ is calculated by $\prod_i \mathcal{L}_i(t_{\text{free}})$ with simulating the Hahn echo signal ($\mathcal{L}_i(t_{\text{free}})$) of isotope $i$, resulting in approximated $T_2$ of a compound by that of isotope $i$ ($T_{2,i}$)

$$T_2 \approx \left(\sum_i T_{2,i}{}^{-\eta_i}\right)^{-\frac{1}{\eta'}}, \tag{1}$$

with $\eta_i$ and $\eta'$ assumed to be 2 in most cases (Methods).

The electron spin $T_{2,i}$ depends mainly on the spin density ($n_i$) of nucleus $i$, the crystalline structure, the nuclear spin *g*-factor ($g_i$), and the nuclear spin quantum number ($I_i$). We compute $T_{2,i}$ with different $n_i$, crystalline structure, and $B$. The nuclear spin density and crystalline structure dependences of $T_{2,i}$ for $^{13}$C is shown in Fig. 2a. For $n_{^{13}\text{C}} < 10^{22}$ cm$^{-3}$ (*cf.* natural abundance in diamond: $1.9\times10^{21}$ cm$^{-3}$), $T_{2,^{13}\text{C}}$ is well fitted by the exponential function $a_{^{13}\text{C}}\, n_{^{13}\text{C}}{}^{-1.0}$, whose exponent −1.0 reproduces previous CCE simulations for diamond and SiC[8,28]. Most importantly, at this density $T_{2,^{13}\text{C}}$ is *independent* of the crystal structure and is governed by interactions between “randomly” positioned $^{13}C$ nucleus. As $n_{\text{spin}}$ increases above $10^{22}$ cm$^{-3}$, the effect of anisotropy of the dipole-dipole interaction[6,7,29] becomes relevant and $T_2$ deviates from the exponential dependence except for in the amorphous limit.

We can therefore scale $T_{2,i}$ as $a_i\, n_i^{-1.0}$, where the coefficient $a_i$, dependent on $g_i$ and $I_i$, is derived by an exponential fit to the calculated $T_{2,i}$ *vs.* $n_i$ as shown in Fig. 2a. Figure 2b then presents $a_i$ for all stable isotopes computed by CCE and the corresponding $T_{2,i}$ at $n_i$ = 1.0×10$^{20}$ cm$^{-3}$ as a function of $g_i$. The calculated data lines up well with different series of $I_i$. The lines are the exponential fits $a_i = b|g_i|^{\beta}$. Figure 2c summarizes the intercept $b$ *vs.* $I_i$ and the exponent $\beta$ *vs.* $I_i$. For spin-1/2 isotopes, $\alpha$ has been analytically calculated to be −13/8 ≈ −1.63[7], which is shown as a dashed line in Fig. 2c, and is in good agreement with numerically obtained $\beta$ = −1.64(7) within the error bar regardless of the $I_i$. We find $b$ changes with $I_i$ and is fitted by the exponential function $b \propto I_i^{-1.10(3)}$ as shown in the dotted line, which indicates that $T_{2,i}$ can be expressed by using $\left(|g_i| I_i^{0.66}\right)^{-1.6}$. Figure 2d shows $T_{2,i}$ *vs.* $|g_i| I_i^{0.66}$, where all the isotopes of all the elements collapse into one line within the error bars. From fitting with an exponential, we determined the phenomenological expression of $a_i$ for all isotopes as $a_i = c|g_i|^{-1.6} I_i^{-1.1}$ with $c$ being an isotope *independent* constant = 1.5×10$^{18}$ cm$^{-3}$s. We therefore obtain the simple expression for $T_{2,i}$ with scaling factors $g_i$, $I_i$, and $n_i$ (in cm$^{-3}$) as

$$T_{2,i} = 1.5 \times 10^{18}\, |g_i|^{-1.6}\, I_i^{-1.1}\, n_i^{-1.0}\ \text{(s)}. \tag{2}$$

This expression obtained by considering CCE of all stable isotopes, combined with equation (1) enables an instantaneous estimate of the $T_2$ with any host materials without treating defect or bath Hamiltonians for dilute nuclear spin baths. This results in a comprehensive prediction of materials with long $T_2$ without the need for any CCE simulations, even for high $I$, or complex heteronuclear systems.

We have assumed defect centers with electron g-factor $g_e = 2$ and $S$ = 1/2 above, while for $S$

> 1/2 centers, a two-level system can be assigned to a given electron spin transition, acting similarly but not equivalently to $S = 1/2$ under the secular approximation. For $S = 1$, for example, $T_2$ is shown to be ~10% longer than that with $S = 1/2$ through CCE calculations[8]. Using a generalized fictitious spin for the magnetic dipole transition $|m_S^-\rangle \leftrightarrow |m_S^+\rangle$ and recalculating the coherence using CCE, we find an expansion of equation (2) that modifies its constant prefactor $c$ for $S = 1/2$ to 3/2 transitions as shown in Fig. 2e. Dashed lines are the exponential fits $c \propto g_e^{\delta}$, and the exponent $\delta$ is $\approx -0.39$, which is in good agreement with theoretically obtained value for $S = 1/2$ and $I = 1/2$ as $-3/8 \approx -0.38$[7]. This is therefore a universal coherence time holding for all transitions for electron spin centers with a dilute spinful nuclear host (Supplemental Information). This expression also hints towards further possible theoretical work that may unravel the physics behind this universal scaling.

In order to prepare for a wide-scale exploration of coherence times for host materials, we investigate the $T_2$ of every element in the periodic table assuming a natural abundance of isotopes, as shown in Fig. 3 taking the element density ($n_{\mathrm{element}}$) of $1.0\times10^{23}$ cm$^{-3}$ (Supplemental Information). This table provides a unique lens to explore and understand the spin coherence of compounds and how to compose materials. Among the elements that compose solid compounds, only cerium has no effect on $T_2$, because all stable isotopes have $I_i = 0$. In addition, there are 7 elements with longer coherence than carbon, which suggests their allotropes or compounds could yield longer coherence time than that of diamond spin centers.

Finally, we demonstrate a comprehensive prediction of $T_2$ based on equations (1) and (2). We utilize structural information from online databases[25,26] to automate the process, considering 12,847 stable materials with predicted bandgaps larger than 1.0 eV. Table I shows the list of the materials with $T_2 > 10$ ms. Here we assume materials have natural isotopic abundance. We attribute the slight deviations of the values on Table I from a full CCE calculation in Fig. 1b to the error on the exponents in equation (2), the anisotropy of dipole-dipole interaction, and that $\eta$ approximated to be 2, as discussed. However, the calculated difference is small and does not hinder screening materials for quantum coherence. We find that $CeO_2$ has the largest $T_2$ of all investigated at 47 ms, which is virtually the upper limit of $T_2$ for all naturally abundant compounds. Beyond choosing the elements of the host crystal and reducing the dimensionality of the host[28], isotopic purification of the material[13,30,31] can further extend $T_2$ coherence times; however, isotopic purification of certain materials is often cost prohibitive or impossible depending on isotopic species.

Of the compounds considered, there are 27 materials with natural isotopic abundance with coherence times longer than 10 ms, all of which are composed of oxides, sulfides, and sulfates. Fig. 4 shows the types of all 832 materials with $T_2 > 1$ ms. SiC has the longest $T_2$ among non-chalcogenides, and our results point to the exploration of chalcogenide materials for longer $T_2$ times than SiC.

We offer a simple, high-throughput method to predict coherence times for spin defects to screen possible quantum host materials. This is achieved by uncovering universal scaling behavior for spin coherence in solids that depends on the effective coherence times of a compound's constituent isotopes. While we do not fully account for geometric factors such as in 2D materials[28], we have demonstrated the coherence time for bulk materials depends only on the nuclear spin *g*-value, its spin quantum number, and density regardless of crystalline structure of the compound. The predictive power of this expression points to 27 materials with coherence time longer than 10 ms and to oxides or sulfides with Ce, Fe, Ca, and Ni as cations as promising long coherence time hosts. In combination with data mining approaches[32], these results present new potential materials systems with promisingly long coherence times, and pave the way to explore unprecedented and varied functional materials for quantum applications.

## Methods

**Spin Hamiltonian, density matrix and its time evolution.** We consider the spin Hamiltonian $\mathcal{H}$ defined by

$$\mathcal{H} = \mathcal{H}_{\mathrm{S}} + \mathcal{H}_{\mathrm{B}} + \mathcal{H}_{\mathrm{S-B}}, \tag{3}$$

where $\mathcal{H}_{\mathrm{S}}$, $\mathcal{H}_{\mathrm{B}}$ are Hamiltonians for electron spin and nuclear spins, respectively, and $\mathcal{H}_{\mathrm{S-B}}$ indicates electron spin – nuclear spin interaction.[8,29,33]

$$\mathcal{H}_{\mathrm{S}} = -g_{\mathrm{e}}\mu_{\mathrm{B}}BS_z, \tag{4}$$

$$\mathcal{H}_{\mathrm{B}} = -\sum_i g_i\mu_{\mathrm{N}}BI_{z,i} + \mathcal{H}_{\mathrm{n-n}}, \tag{5}$$

$$\begin{aligned}\mathcal{H}_{\mathrm{S-B}} &\approx \frac{\mu_0}{4\pi} g_{\mathrm{e}}\mu_{\mathrm{B}}\mu_{\mathrm{N}}\vec{S}\cdot\sum_i g_i\left[\frac{\vec{I}_i}{{r_i}^3} - \frac{3(\vec{I}_i\cdot\vec{r}_i)\vec{r}_i}{{r_i}^5}\right] \\ &\approx S_z\sum_i \vec{A}_i\cdot\vec{I}_i,\end{aligned} \tag{6}$$

where $g_{\mathrm{e}}$, $g_i$, $\mu_{\mathrm{B}}$, $\mu_{\mathrm{N}}$, and $\mu_0$ are the *g*-factor of electron, the *g*-factor of nuclear spin of nucleus $i$, Bohr magneton, nuclear magneton, and the permeability of vacuum, respectively. We set the magnetic field direction to be $z$ direction, and electron spin quantum number to be 1/2. $\vec{r}_i$, $r_i$, $B$, $\vec{A}_i$, $\vec{S}$, $S_z$, $\vec{I}_i$, and $I_{z,i}$ are vectors from electron spin to the nucleus $i$, $|\vec{r}_i|$, the magnetic field, hyperfine field vector of nucleus $i$, the electron spin vector operator, $z$ component of electron spin operator, the spin operator of nucleus $i$, and $z$ component of spin operator of nucleus $i$, respectively. $\mathcal{H}_{\mathrm{n-n}}$ is the Hamiltonian of nuclear spin – nuclear spin interactions;

$$\mathcal{H}_{\mathrm{n-n}} = \frac{\mu_0}{4\pi}\mu_{\mathrm{N}}^2\sum_{\{i,j\}} g_i g_j\left[\frac{\vec{I}_i\cdot\vec{I}_j}{r_{ij}^3} - \frac{3(\vec{I}_i\cdot\vec{r}_{ij})(\vec{I}_j\cdot\vec{r}_{ij})}{r_{ij}^5}\right], \tag{7}$$

where $\vec{r}_{ij}$ is the vector from nucleus $i$ to nucleus $j$, and $r_{ij} = |\vec{r}_{ij}|$. Two of the approximations in equation (6) are valid when (1) Fermi contact term is negligible with localized electron spin center

and dilute nuclear spin in the host, which are valid in the most of the intrinsic and extrinsic defects in, *e.g.*, SiC and diamond, and (2) two of the electron spin states $m_\mathrm{S} = \pm 1/2$ are of order GHz, *e.g.*, when one applies, for the $g = 2$ defect, a magnetic field larger than 30 mT, which is standard measurement condition for pseudospin model, respectively.

Time evolution of the density matrix $\rho(t_\mathrm{free})$ is calculated by

$$\rho(t_\mathrm{free}) = \mathcal{U}(t_\mathrm{free})\rho(0)\mathcal{U}^\dagger(t_\mathrm{free}). \tag{8}$$

We use the standard Hahn echo propagator composed of $(\pi/2)_x$ pulse, free evolution for $t_\mathrm{free}/2$, $\pi_x$ pulse, and free evolution for $t_\mathrm{free}/2$, as

$$\mathcal{U}(t) = \exp\left(-i\frac{\mathcal{H}}{\hbar}\frac{t_\mathrm{free}}{2}\right)\exp(i\pi S_x)\exp\left(-i\frac{\mathcal{H}}{\hbar}\frac{t_\mathrm{free}}{2}\right)\exp\left(i\frac{\pi}{2}S_x\right). \tag{9}$$

The initial density matrix is taken to be $\rho(0) = \rho_\mathrm{S}(0)\otimes\rho_\mathrm{B}(0)$ using electron spin projected density matrix $\rho_\mathrm{S}(0)$ with $z$ projection of spin $m_\mathrm{S} = -1/2$ state

$$\rho_\mathrm{S}(0) = \left|-\frac{1}{2}\right\rangle\left\langle-\frac{1}{2}\right|, \tag{10}$$

and bath projected density matrix $\rho_\mathrm{B}(0)$

$$\rho_\mathrm{B}(0) = \sum_{\mathcal{J}} \mathcal{P}_{\mathcal{J}}|\mathcal{J}\rangle\langle\mathcal{J}| \tag{11}$$

with $\mathcal{P}_{\mathcal{J}}$ being the probability of the nuclear state $|\mathcal{J}\rangle$. Hahn echo signal $\mathcal{L}(t_\mathrm{free})$ is calculated by

$$\mathcal{L}(t) = \frac{\mathrm{Tr}[\rho(t_\mathrm{free})S_+]}{\mathrm{Tr}[\rho(0)S_+]}, \tag{12}$$

where $S_+$ is raising operator of electron spin[21].

**Cluster correlation expansion (CCE) calculation.** Hahn echo signal $\mathcal{L}^{\mathrm{CCE}-1}$ obtained by first-, and second-order CCE (CCE-1, and CCE-2) calculations, respectively, are defined as[3]

$$\mathcal{L}^{\mathrm{CCE-1}} = \prod_i \mathcal{L}_i, \quad (13)$$

$$\mathcal{L}^{\mathrm{CCE-2}} = \mathcal{L}^{\mathrm{CCE-1}} \prod_{\{i,j\}} \frac{\mathcal{L}_{i,j}}{\mathcal{L}_i \mathcal{L}_j}, \quad (14)$$

where $\mathcal{L}_i$ ($\mathcal{L}_{i,j}$) is the Hahn echo signals calculated with the central electron spin and the $i$-th nuclear spin (electron spin and the $i$-th and $j$-th nuclear spins). It is known that in the dilute nuclear spin bath like SiC, the effect of the three or higher body spin interaction is rare and the $\mathcal{L}(t_{\mathrm{free}})$ converges with second-order CCE[8,28].

**Decoupling field.** The envelope of the Hahn-echo signal is critically affected by the dipole-dipole interactions between nuclear spins. The dipole-dipole interaction between heteronuclear spin is characterized by two factors: $\Omega$ and $\delta$. $\Omega$ indicates the dipole-dipole interactions between nucleus $i$ and $j$, which is given by equation (7). $\delta$ indicates the energy splitting between two levels interacting with $I_{+,i}I_{-,j} + I_{-,i}I_{+,j}$ due to the different Zeeman splitting with different nuclear spin $g$-factors between nucleus in addition to the dipole-dipole interaction between them with $I_{\pm,i}$ being the ladder operator of spin in nucleus $i$ given by equation (5). When $\delta \gg \Omega$, the heteronuclear spin baths are decoupled. Considering $I_{+,i}I_{-,j} + I_{-,i}I_{+,j}$ is the main source of the decoherence[8], we estimate decoupling field $B_{\mathrm{dec}}$ as

$$B_{\mathrm{dec}} = \frac{\mu_0}{4\pi} \mu_{\mathrm{N}} \frac{1}{l^3} \frac{g_i g_j}{g_i - g_j}, \quad (15)$$

with $l$ being the distance of the nearest-neighbor nucleus $i$, and nucleus $j$, respectively (Supplemental Information). For example, $B_{\mathrm{dec}}$ is 0.28 mT (0.13 mT) for $SiO_2$ (SiC), above which the heteronuclear spin baths decouple[34].

Using CCE calculations, Seo *et al*., have numerically shown that $B$ < 30 mT decouples heteronuclear spin baths assuming the difference of nuclear spin $g$-factor values ($\Delta g$) = 0.021 and $l$ = 1.3 Å[8]. These $\Delta g$ and $l$ values are relatively small among the compounds. Also, in experiments, $B$ up to 300 mT ~ 1 T is achievable with a standard yoke magnet. In equation (15), the decoupling field $B_{\mathrm{dec}}$ is proportional to $1/l^3\Delta g$, thus, suggesting the heteronuclear spin baths are decoupled in the most of the compounds under standard experimental conditions.

As example systems, let us consider the oxide and sulfides. The ionic radius of the $O^{2-}$ is 0.14 nm at minimum, thus $B_{\mathrm{dec}}$ is estimated to be ~$g_{\mathrm{O}}^2/\Delta g$ × 0.9 mT at most by equation (15), with $\Delta g$ being the difference of the $g$-factors between $^{17}O$ and cation. For the worst case among all isotopes, $\Delta g$ = 0.024 for $^9Be$ gives a maximum $B_{\mathrm{dec}}$ ~5 mT. For sulfides the largest $B_{\mathrm{dec}}$ is given by $^{189}Os$ with $\Delta g$ = 0.011, as ~3 mT.

**Stretching exponent.** A compound's $T_2$ is defined by each isotope's coherence time ($T_{2,i}$) by the condition $\sum_i (T_2/T_{2,i})^{-\eta_i} = 1$, where $\eta_i$ is the stretching exponent for the $\mathcal{L}_i(t)$. We find this $T_2$ is well approximated by

$$T_2 \approx \left(\sum_i T_{2,i}{}^{-\eta_i}\right)^{-\frac{1}{\eta'}}, \tag{16}$$

with $\eta_i$ and $\eta'$ assuming to be 2. For an example, when $T_{2,j} = T_{2,i}/10$ ($T_{2,j} = T_{2,i}/3$), $T_2$ in binary compound with nucleus $i$ and $j$ obtained by equation (1) with $\eta_i = \eta'$ = 2 deviates from the exact $T_2$ by 0.44% (4.0%) at the very most among the typical $\eta_i$ and $\eta_j$ values 2–3[8].

**Materials explorations.** For $T_2$ prediction, we have used crystallographic information framework (CIF) files available at The Materials Project[25,26]. From CIF files, $n_i$ is derived and $T_2$ is calculated by using equations (1) and (2). Only the predicted but realistic and stable materials *i.e.*, materials with zero-energy above hull are calculated.

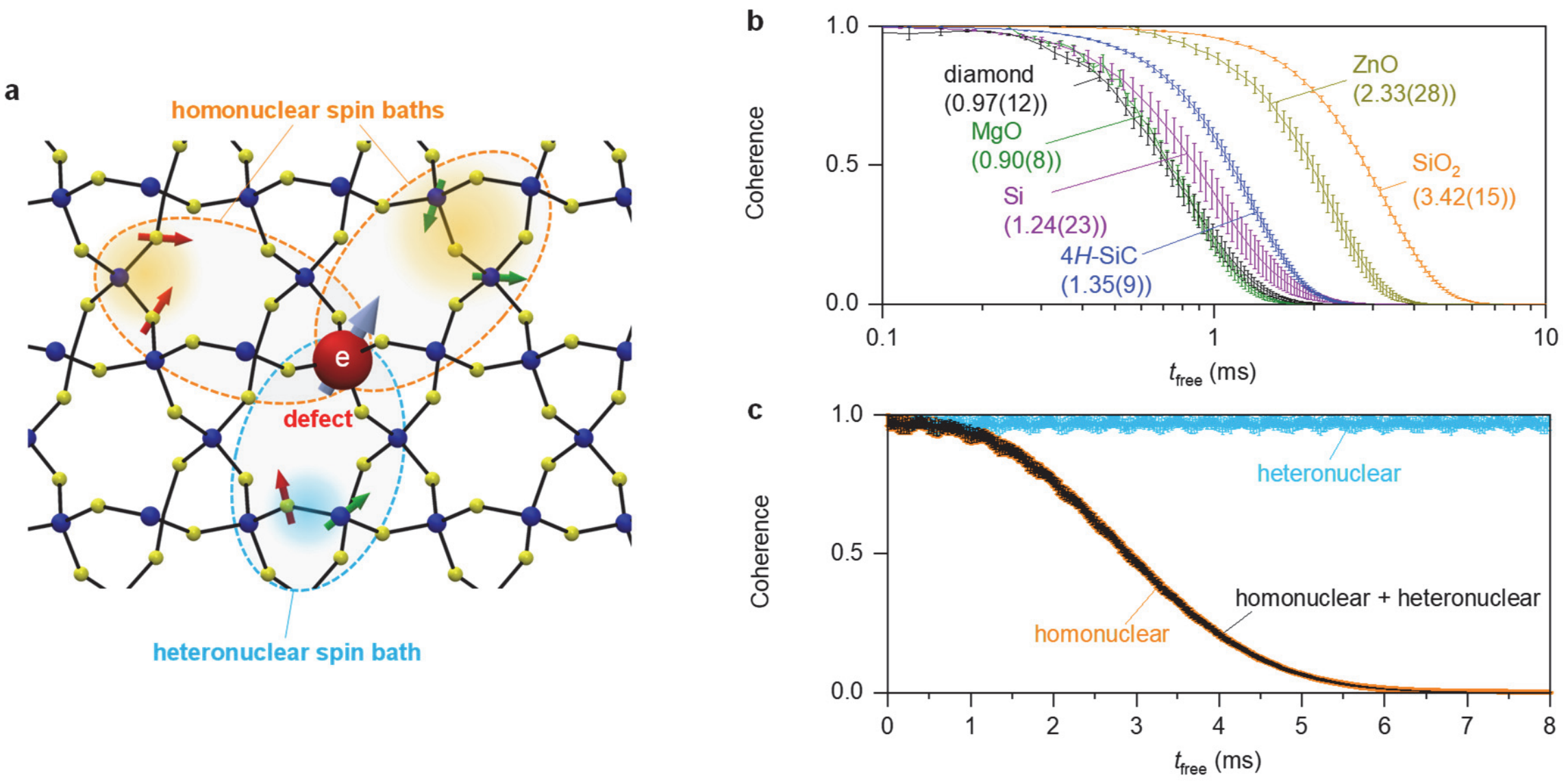

**Fig. 1 | Quantum spin coherence simulation. a**, Schematics of second-order cluster correlation expansion (CCE-2) of a defect electron spin in a heteronuclear compound. **b,** Hahn echo signal $\mathcal{L}(t_{\mathrm{free}})$ *vs*. free evolution time $t_{\mathrm{free}}$ calculated by CCE-2 for naturally-abundant isotopic diamond, 4*H*-SiC, silicon, and oxides obtained by simulation under external magnetic field $B$ of 5 T. **c**, $\mathcal{L}(t_{\mathrm{free}})$ of $SiO_2$ ($\alpha$-quartz) with $B$ = 300 mT. In addition to the $\mathcal{L}(t_{\mathrm{free}})$ with dipole-dipole interactions with all baths (black), that with solely homonuclear spin bath (orange), and hetero-nuclear spins (blue) are shown. Error bars indicate the sample standard deviation of the Hahn echo signal for different instances of nuclear spin coordinates.

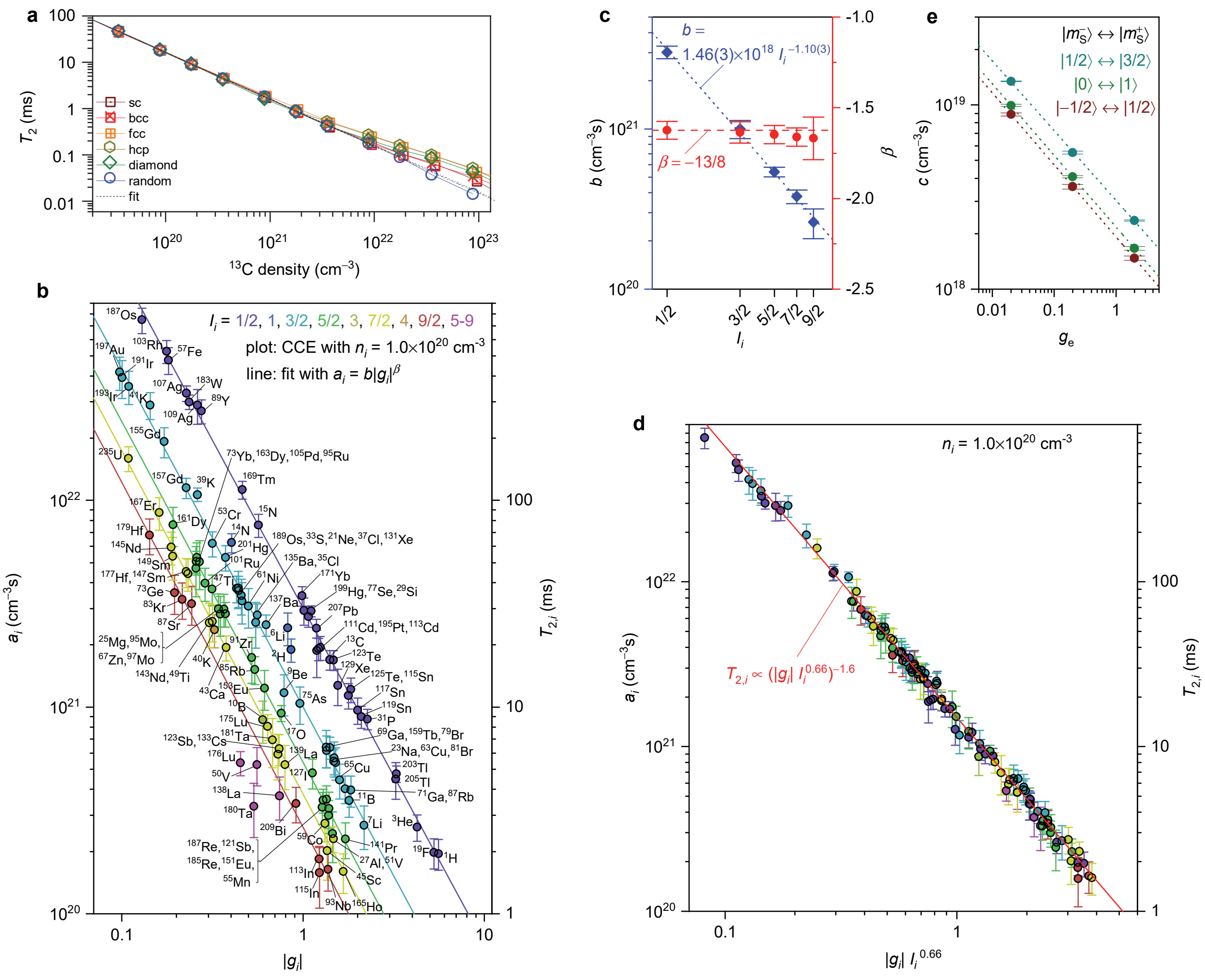

**Fig. 2 | Scaling of quantum coherence of decoupled spin baths. a**, Predicted quantum coherence time $T_2$ of defects in crystals composed of carbon as a function of $^{13}$C density $n_i(i = {}^{13}\mathrm{C})$ with various crystal structures. Dashed line shows the fit with an exponential function $a_i n_i^{\alpha}$, with $a_i$ being coefficient, $\alpha$ the exponent −1.0. An external magnetic field of 5 T is applied along [111] direction of diamond structure and along [001] directions of other crystal structures. **b,** Coefficient $a_i$ and corresponding $T_2$ with nuclear spin density $n_i$ = 1.0×10$^{20}$ cm$^{-3}$ as a function of the absolute value of nuclear spin *g*-factor $|g_i|$ calculated for all stable isotopes with the nuclear spin quantum number $I_i$. Lines are exponential fits $T_{2,i} \propto b|g_i|^{\beta}$ on the different half-integer-$I_i$ spins. Error bars indicate the sample standard deviation obtained by the simulation for different crystal coordinates for the isotopes. **c**, Intercept $b$ *vs.* $I_i$ with the exponential fit $b = cI_i^{-1.10(3)}$ (blue) with $c$ being the coefficient, and the exponent $\beta$ *vs.* $I_i$ with the theoretical value $\beta$ = −13/8 for $I_i$ = 1/2[7,14] (red). **d**, $T_2$ *vs.* $|g_i|{I_i}^{0.66}$. Solid line is the exponential fit. All simulations are conducted under external magnetic field of 5 T. **a**–**d**, Electron *g*-factor $g_e$ = 2.0 and $S$ = 1/2 are assumed. **e**, Coefficient $c$ for the transition

of electron spin states between $|m_S^-\rangle \leftrightarrow |m_S^+\rangle$ as a function of $g_e$. Dashed lines are the exponential fits. Error bars indicate the standard error obtained from fitting the simulated CCE data.

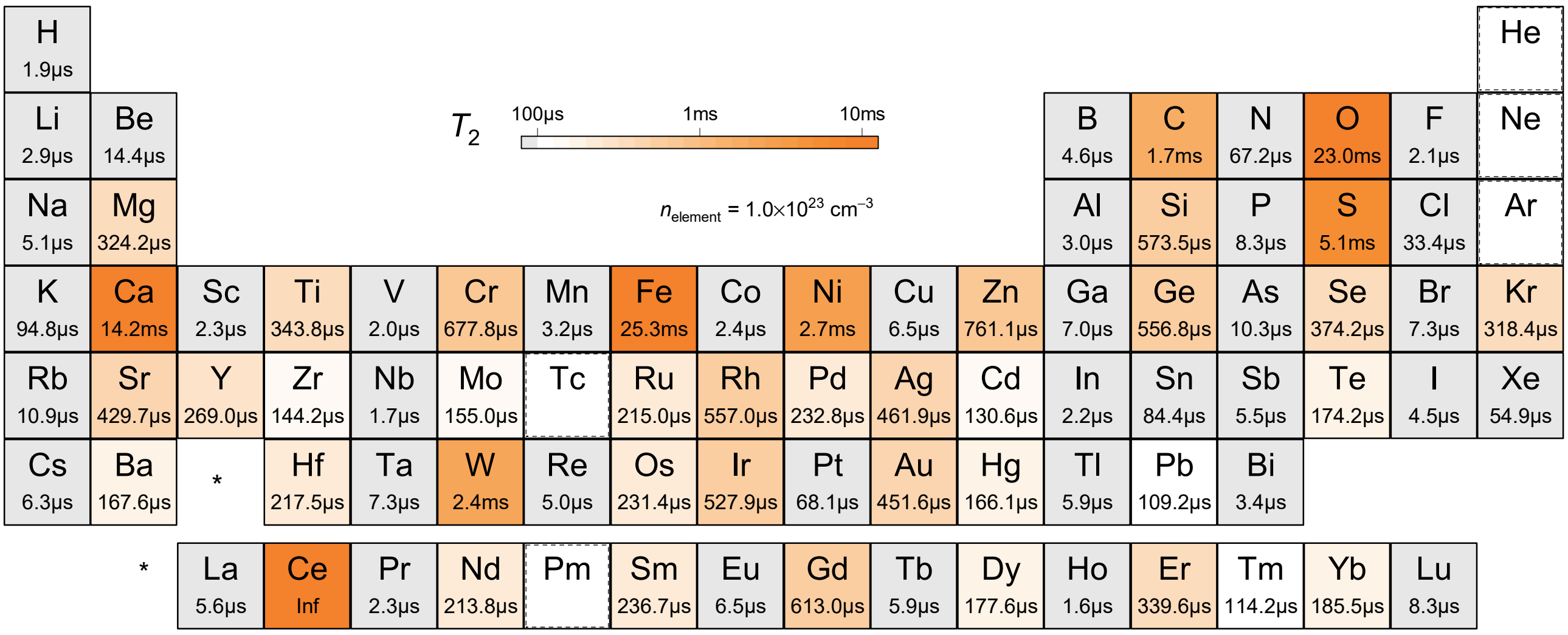

**Fig. 3 | Periodic table for quantum coherence.** Coherence time $T_2$ simulated with CCE calculations for spin qubits in hypothetical material hosts with natural abundance of a single species with element density $n_{\mathrm{element}}$ = $1.0\times10^{23}$ cm$^{-3}$. The periodic table is 'painted' by $T_2$ on a log scale. Materials which are difficult to make compounds from (He, Ne, Ar) or that are without stable isotopes (Tc, Pm) are excluded.

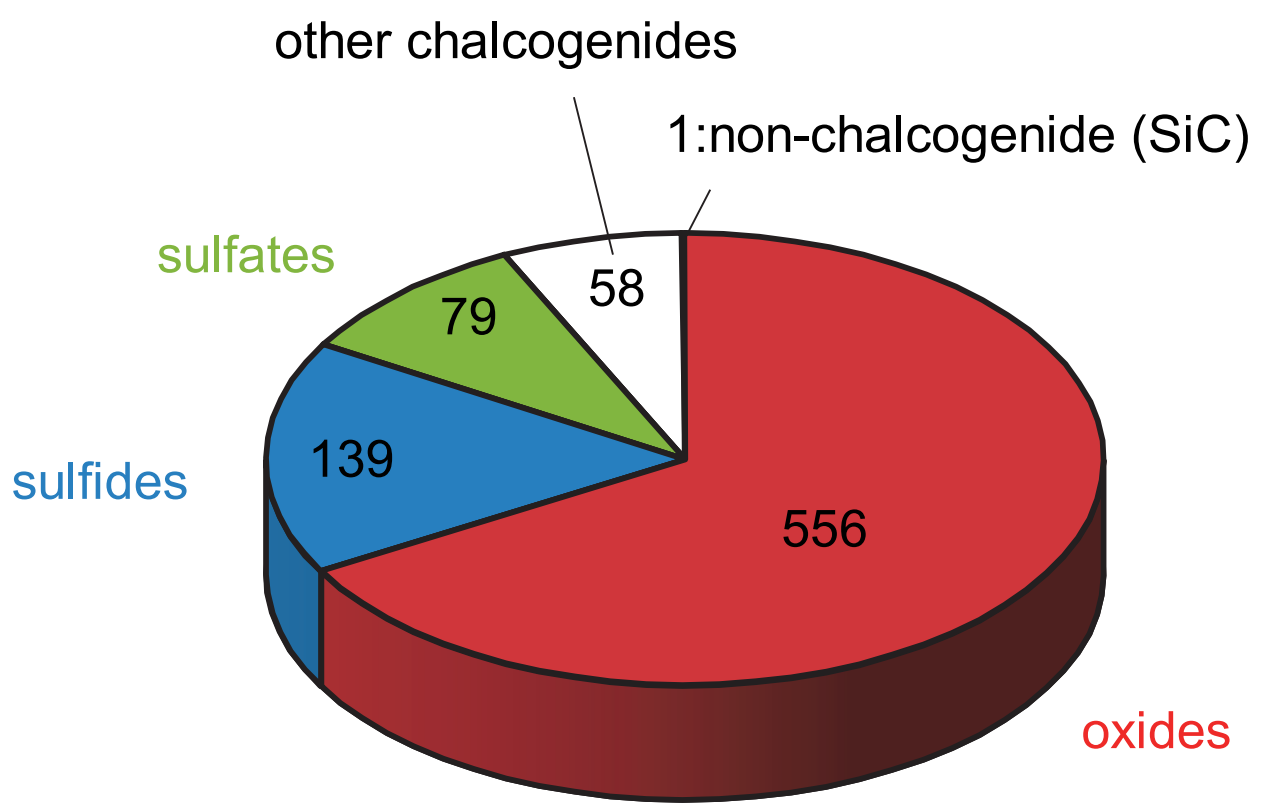

**Fig. 4 | Materials to explore.** Types of 832 stable compounds with quantum coherence time $T_2$ longer than 1 ms and bandgap larger than 1.0 eV. SiC is the only stable widegap non-chalcogenide with $T_2$ > 1 ms.

**Table 1 | Top quantum coherence time $T_2$ materials**. Materials with $T_2$ > 10 ms and bandgap > 1 eV, as well as those listed in Fig. 1b are shown.

| # | Material | $T_2$ (ms) | # | Material | $T_2$ (ms) | # | Material | $T_2$ (ms) |
|---|---|---|---|---|---|---|---|---|
| 1 | $CeO_2$ | 47 | 12 | CaS | 23 | 23 | $WS_2$ | 11 |
| 2 | FeO | 36 | 13 | $Ca_2NiWO_6$ | 19 | 24 | $Sr_2Si(S_2O_7)_4$ | 11 |
| 3 | CaO | 34 | 14 | S | 19 | 25 | $Sr_2Ge(S_2O_7)_4$ | 11 |
| 4 | $CaSO_4$ | 29 | 15 | $CaWO_4$ | 18 | 26 | $CaCO_3$ | 11 |
| 5 | $Ce(SO_4)_2$ | 29 | 16 | $CS_{14}$ | 18 | 27 | $FeS_2$ | 10 |
| 6 | $SO_3$ | 29 | 17 | $Fe_2NiO_4$ | 18 | | | |
| 7 | $FeSO_4$ | 28 | 18 | $S_8O$ | 17 | 138 | $SiO_2$ | 2.7 |
| 8 | $CaS_3O_{10}$ | 28 | 19 | $FeWO_4$ | 16 | 298 | ZnO | 1.9 |
| 9 | $Ca_3WO_6$ | 27 | 20 | $NiSO_4$ | 15 | 709 | SiC | 1.1 |
| 10 | $WS_2O_9$ | 25 | 21 | $WO_3$ | 13 | 936 | diamond | 0.89 |
| 11 | $Ca_2FeWO_6$ | 24 | 22 | $NiWO_4$ | 12 | 1125 | MgO | 0.60 |

## Acknowledgements

We thank Tomasz Dietl, Fumihiro Matsukura, William A. Borders, Shunsuke Fukami, and Nikita Onizhuk for fruitful discussions, He Ma, Jaewook Lee, and Huijin Park for their help in cross-checking the CCE predictions. This work was supported in part by Marubun Research Promotion Foundation, RIEC through Overseas Training Program for Young Profession and Cooperative Research Projects, MEXT through the Program for Promoting the Enhancement of Research Universities, JSPS Kakenhi Nos. 19KK0130 and 20H02178, the EFRC Center for Novel Pathways to Quantum Coherence in Materials (NPQC) supported by DOE/BES, National Research Foundation of Korea (NRF) grant funded by the Korea government (MSIT) (No. 2018R1C1B6008980, No. 2018R1A4A1024157, and No. 2019M3E4A1078666), AFOSR FA9550-19-1-0358. Work at Argonne was supported primarily by the Center for Novel Pathways to Quantum Coherence in Materials, an Energy Frontier Research Center funded by the U.S. Department of Energy, Office of Science, Basic Energy Sciences in collaboration with the U.S. Department of Energy, Office of Science, National Quantum Information Science Research Centers.

## Author contributions

D.D.A, H.O., F.J.H., and S.K. planned the study. S.K. performed CCE calculations with input from H.S., F.J.H., G.W., C.P.A., and S.E.S. All authors discussed the results and contributed to the preparation of the manuscript.

## Competing interests

The authors declare no competing interests.

### Data availability

The datasets generated and analyzed during the current study are available from the corresponding author upon reasonable request.